\documentclass[iop]{emulateapj}

\newcommand{\dgr}{^{\circ}}
\newcommand{\vb}{{\bf b}}
\newcommand{\vp}{w_{\parallel}}
\newcommand{\vg}{w_g}
\newcommand{\vn}{{\bf n}}
\newcommand{\vv}{{\bf v}}
\newcommand{\vw}{{\bf w}}
\newcommand{\vz}{\boldsymbol{\zeta}}
\newcommand{\vx}{\boldsymbol{\xi}}
\newcommand{\cth}{\cos\vartheta}
\newcommand{\sth}{\sin\vartheta}
\newcommand{\thl}{\vartheta_{\rm loss}}
\renewcommand{\theta}{\vartheta}

\usepackage[colorlinks,urlcolor=blue,citecolor=blue,linkcolor=blue]{hyperref}
\usepackage{amsmath}

\shorttitle{Injection at Shocks}
\shortauthors{Caprioli, Pop, Spitkovsky}

\begin{document}

\title{Simulations and Theory of Ion Injection at Non-relativistic Collisionless Shocks}

\author{Damiano Caprioli, Ana-Roxana Pop, Anatoly Spitkovsky}
\affil{Department of Astrophysical Sciences, Princeton University, 
    4 Ivy Ln., Princeton NJ 08544, USA}
\email{caprioli@astro.princeton.edu}

\begin{abstract}
We use kinetic hybrid simulations (kinetic ions -- fluid electrons) to characterize the fraction of ions that are accelerated to non-thermal energies at non-relativistic collisionless shocks. We investigate the properties of the shock discontinuity and show that shocks propagating almost along the background magnetic field (quasi-parallel shocks) reform quasi-periodically on ion cyclotron scales. Ions that impinge on the shock when the discontinuity is the steepest are specularly reflected. This is a necessary condition for being injected, but it is not sufficient. Also by following the trajectories of reflected ions, we calculate the minimum energy needed for injection into diffusive shock acceleration, as a function of the shock inclination. We construct a minimal model that accounts for the ion reflection from quasi-periodic shock barrier, for the fraction of injected ions, and for the ion spectrum throughout the transition from thermal to non-thermal energies. This model captures the physics relevant for ion injection at non-relativistic astrophysical shocks with arbitrary strengths and magnetic inclinations, and represents a crucial ingredient for understanding the diffusive shock acceleration of cosmic rays.	
\end{abstract}


\section{Introduction}
Diffusive shock acceleration \cite[DSA; e.g.,][]{bell78a,blandford-ostriker78} at non-relativistic collisionless shocks is a prominent mechanism for producing very energetic particles.
It is particularly efficient at supernova remnant (SNR) blast waves \citep[e.g.,][]{tycho}, and is likely responsible for the acceleration of Galactic cosmic rays (CRs).
Nevertheless, determining the exact fraction of particles that are injected into DSA is a vexed question in CR physics.
A characterization of particle injection without free parameters requires a self-consistent calculation of the shock structure on microphysical scales, which can be achieved only with kinetic plasma simulations.

In this paper we develop a simplified predictive model of injection that matches results from hybrid simulations (kinetic ions--fluid electrons).
From simulations we infer the dynamics of the shock transition (Section \ref{sec:reform}) and characterize the trajectories of accelerated ions after their first shock reflection, determining the conditions for injection into DSA as a function of the magnetic field orientation (Section \ref{sec:inj}).
We then use these relations to construct a minimal theory of ion injection at DSA-efficient shocks, which accounts for the fraction of accelerated ions, and for the ion spectrum from thermal to non-thermal energies  (Section \ref{sec:suprath}).
We discuss our results in Section \ref{sec:disc}, and conclude in Section \ref{sec:conclusions}.   
\begin{figure}\centering
\includegraphics[trim=0px 50px 0px 240px, clip=true, width=.485\textwidth]{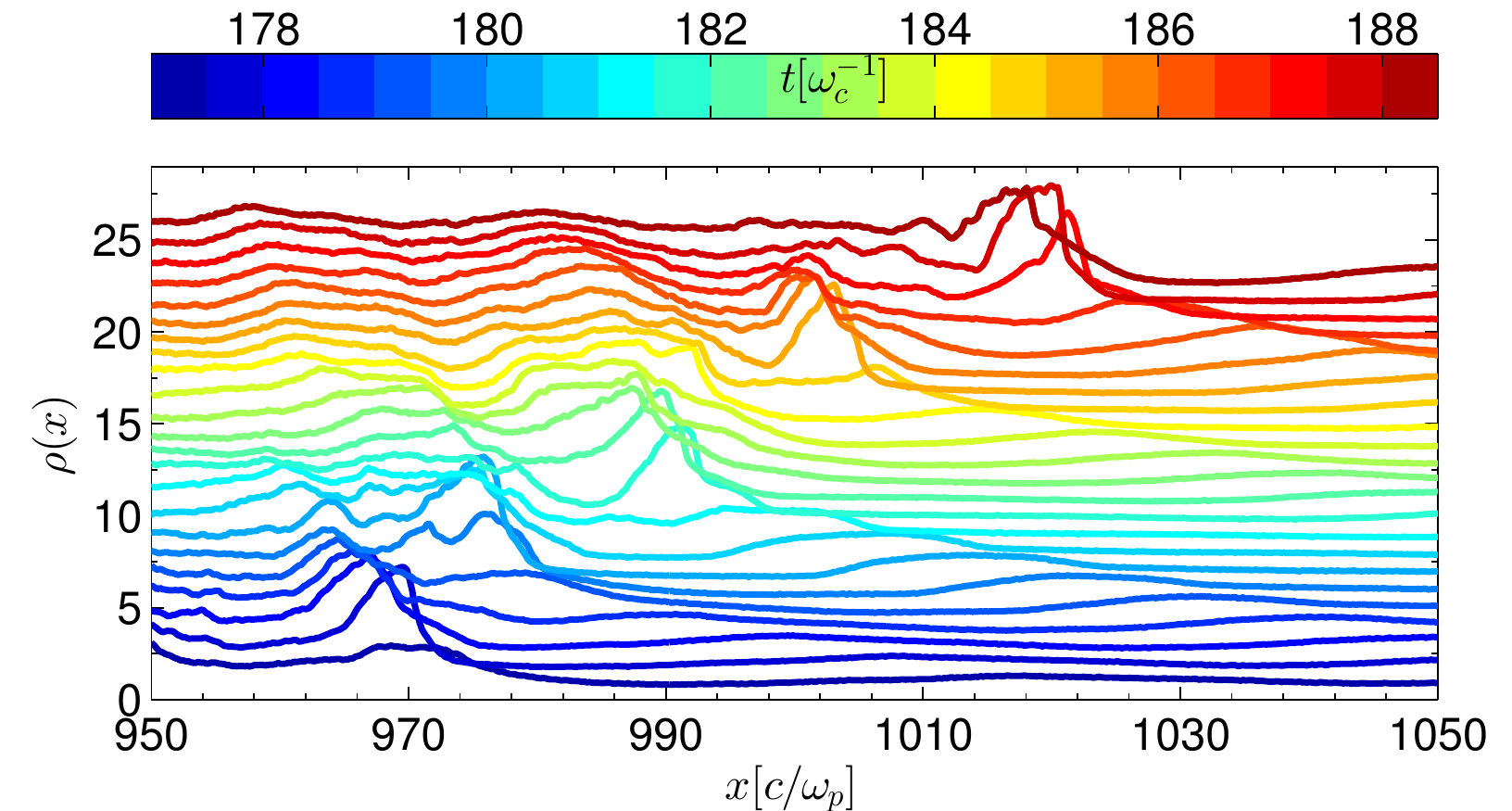}
\includegraphics[trim=0px 50px 0px 280px, clip=true, width=.485\textwidth]{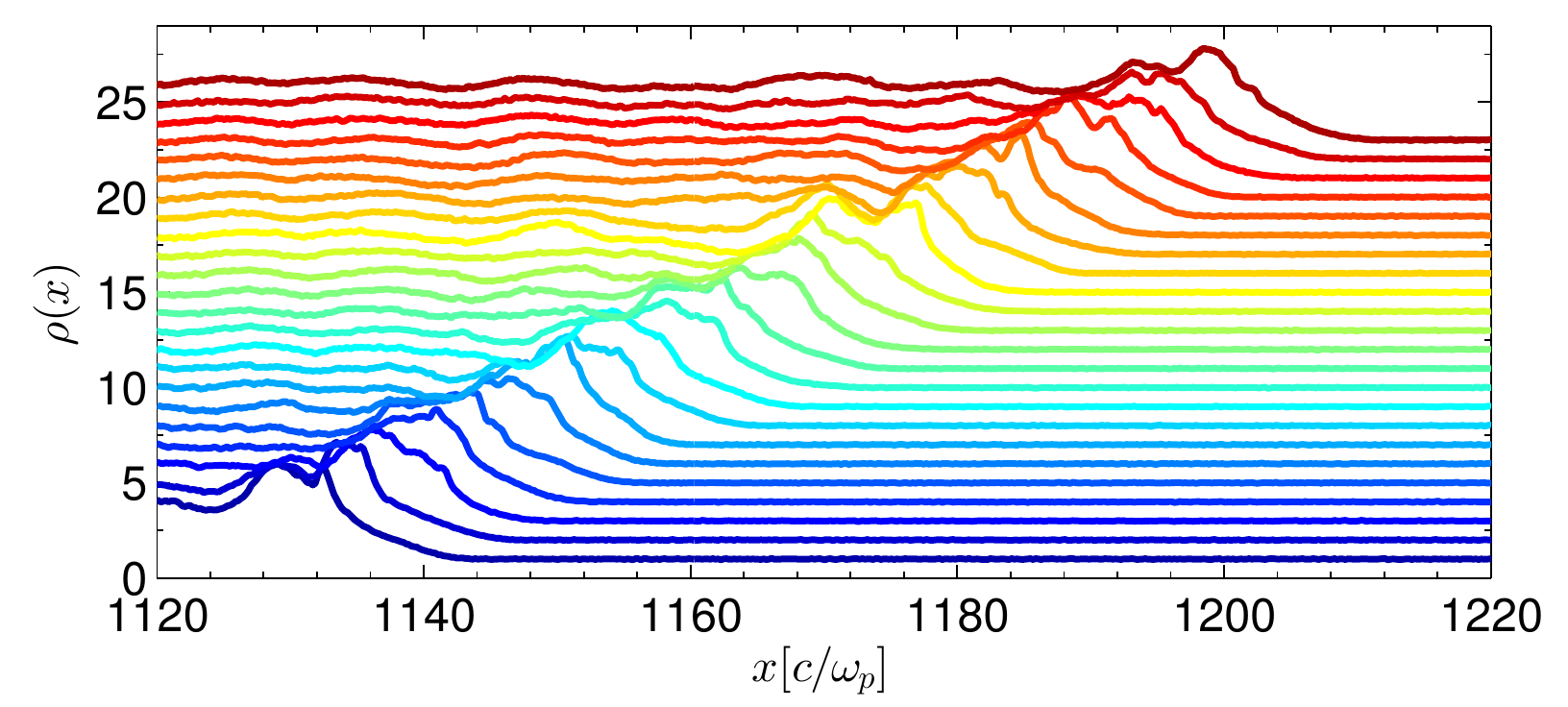}
\caption{\label{fig:rho} Evolution of the density profile for a parallel shock ($\theta=0\dgr$, top panel), and a quasi-perpendicular shock ($\theta=80\dgr$, bottom) with $M=20$. 
For better readability, profiles at later times are shown as increasingly shifted up. 
}
\end{figure}

\section{Hybrid simulations}\label{sec:reform}
Hybrid simulations describe the evolution of a collisionless plasma integrated over the electron scales, fully retaining the physics of shock formation, ion acceleration, and magnetic field amplification, as we discussed in a recent series of papers \citep[][hereafter Paper I, II, III]{DSA,MFA,diff}.
Here we present 2D simulations performed with the massively-parallel code \emph{dHybrid} \citep{gargate+07}, where a shock is produced by sending a supersonic fluid against a reflecting wall (see Paper I). 
Lengths are in units of $c/\omega_p$, where $c$ is the speed of light and $\omega_p\equiv\sqrt{4\pi n e^2/m}$, with $m,e$, and $n$ the ion mass, charge, and number density.
Time is in units of $\omega_c^{-1}\equiv mc/eB_0$, with ${\bf B}_0\equiv B_0{\bf b}$ the background magnetic field; 
the time step is $\Delta t=0.003\omega_c^{-1}$.
Velocities and energies are normalized to $v_A\equiv B_0/\sqrt{4\pi m n}$ and $E_{\rm sh}\equiv m(M_Av_A)^2/2$.
Sonic and Alfv\'enic Mach numbers are $M_s\approx M_A\equiv v_{sh}/v_A\equiv M$, with ${\bf v}_{\rm sh}=-v_{\rm sh}{\bf x}$ the upstream fluid velocity in the simulation frame.
Boxes measure $30,000c/\omega_p\times 40c/\omega_p$, with 5 cells per $c/\omega_p$ and 4 particles per cell. 
The shock inclination is expressed by $\theta\equiv \arccos({\bf b}\cdot {\bf x})$.

\begin{figure}\centering
\includegraphics[trim=25px 230px 25px 205px, clip=true, width=.485\textwidth]{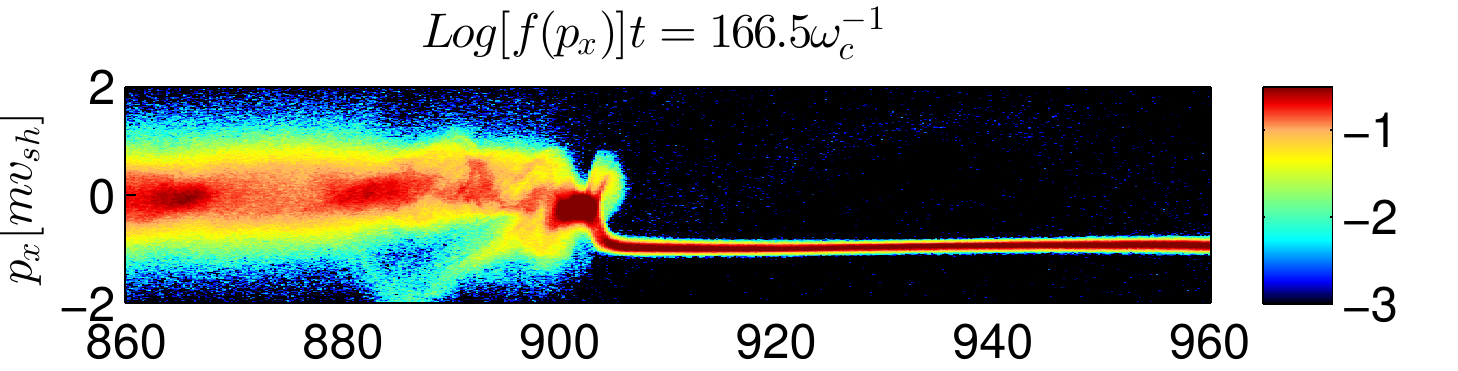}
\includegraphics[trim=25px 230px 25px 205px, clip=true, width=.485\textwidth]{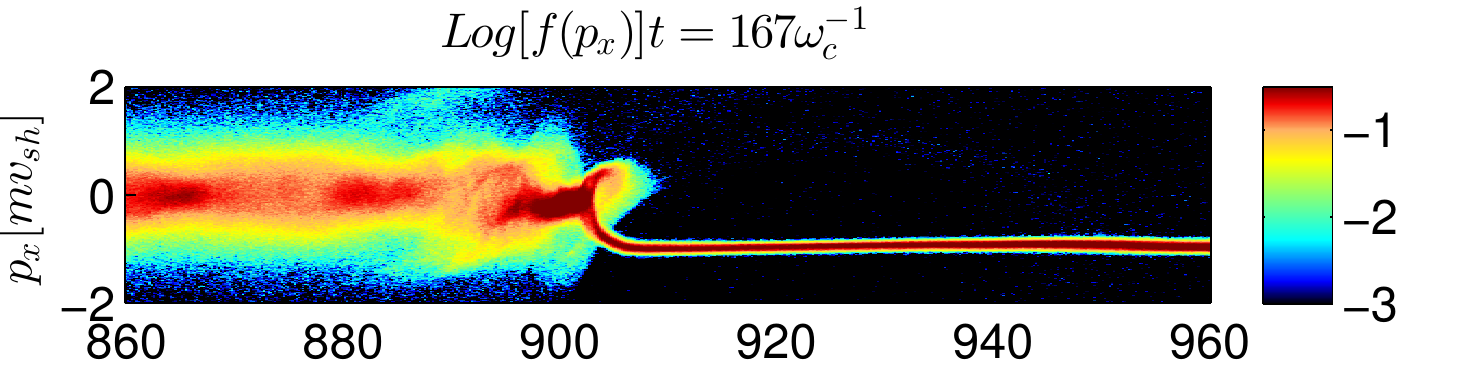}
\includegraphics[trim=25px 230px 25px 205px, clip=true, width=.485\textwidth]{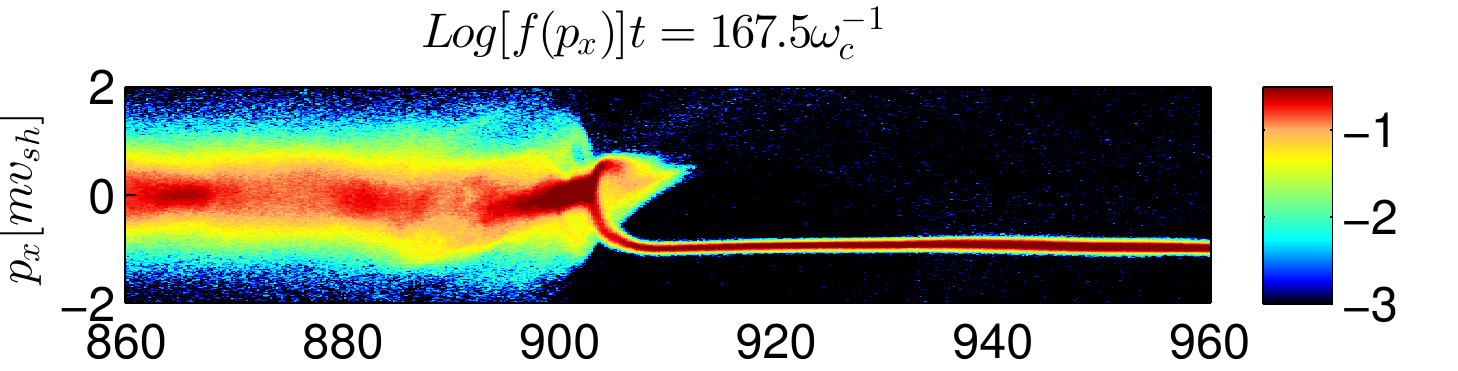}
\includegraphics[trim=25px 230px 25px 205px, clip=true, width=.485\textwidth]{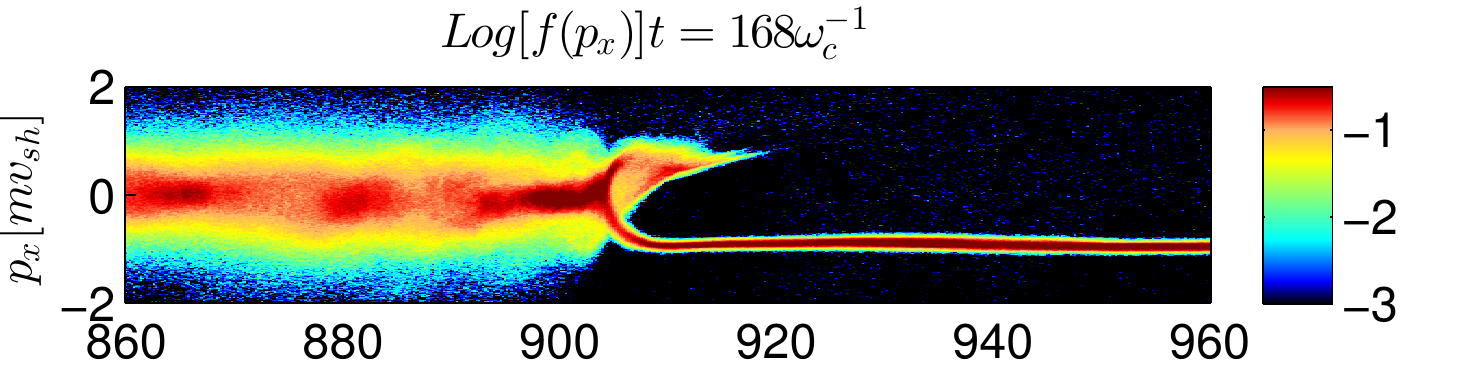}
\includegraphics[trim=25px 230px 25px 205px, clip=true, width=.485\textwidth]{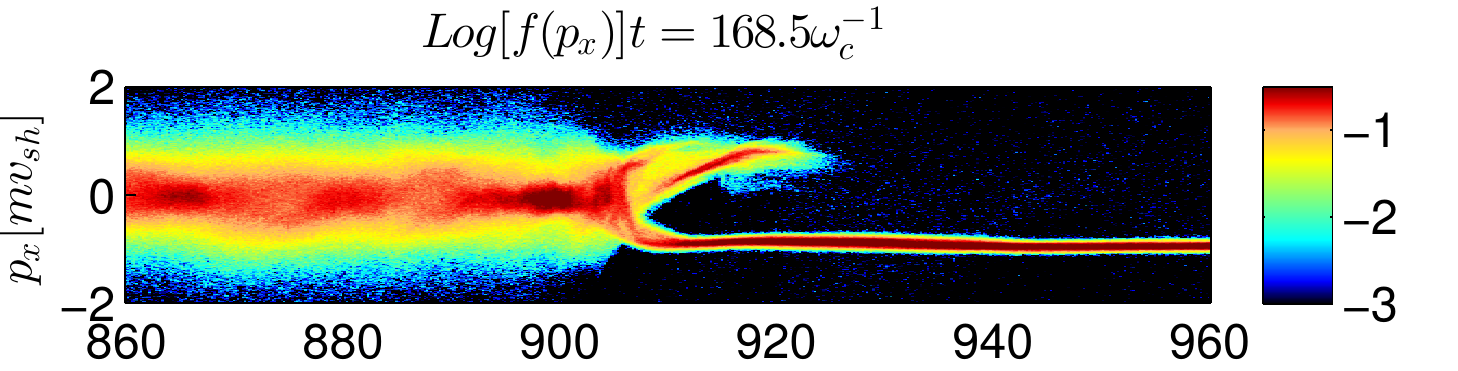}
\includegraphics[trim=25px 230px 25px 205px, clip=true, width=.485\textwidth]{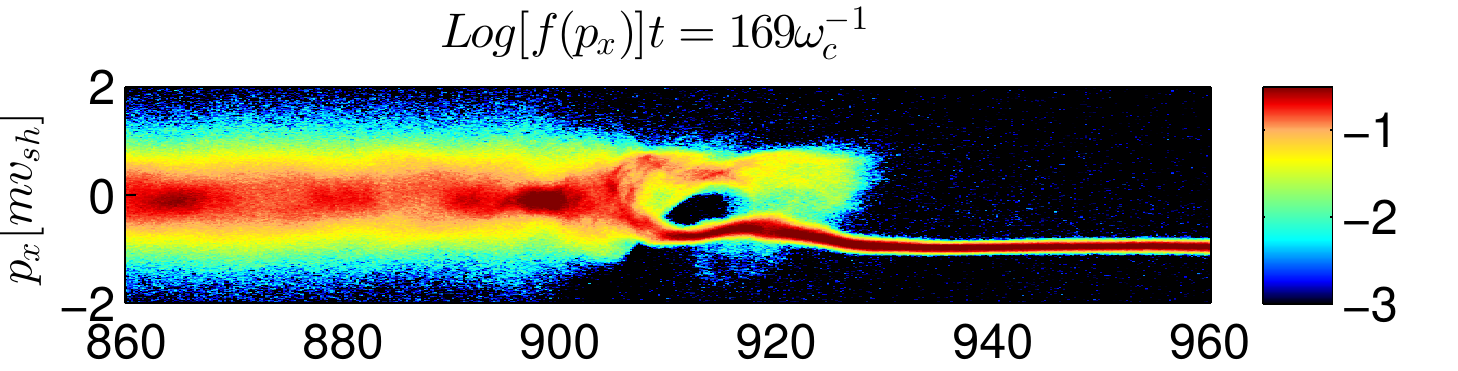}
\includegraphics[trim=25px 210px 25px 205px, clip=true, width=.485\textwidth]{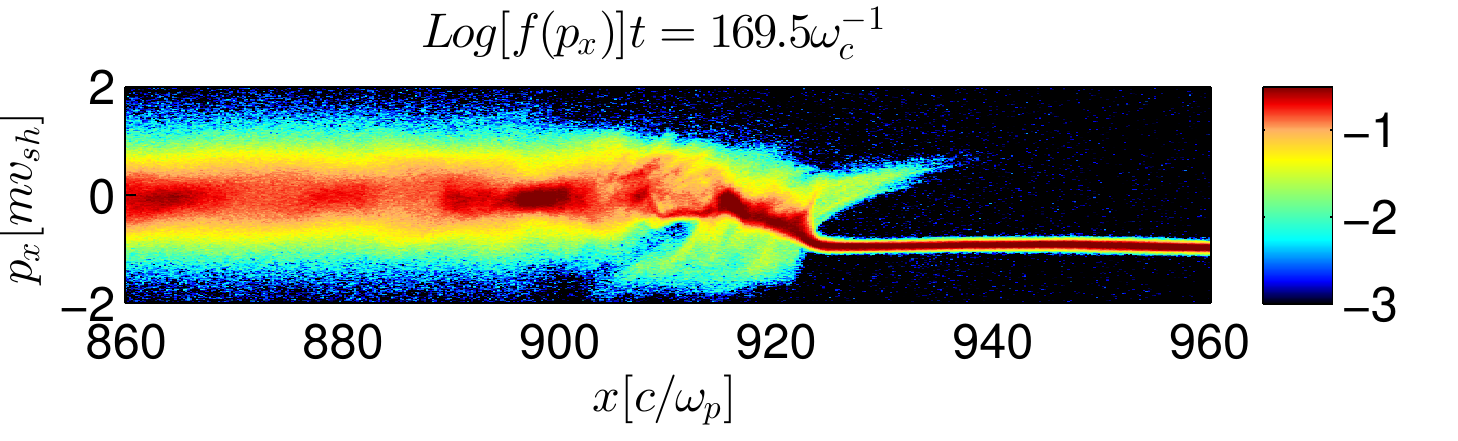}
\caption{\label{fig:trans} Evolution of the $x-p_x$ phase space distribution for a parallel shock with $M=20$.
The discontinuity is almost steady for $\sim 3\omega_c^{-1}$, until the beam of reflected ions induces its reformation.}
\end{figure}

Figure \ref{fig:rho} shows the evolution of the density profile for a parallel ($\theta=0\dgr$) and a quasi-perpendicular ($\theta=80\dgr$) shock;
in both cases $M=20$ and the discontinuity propagates with the same average speed.
For $\theta=0$ the density peak first broadens and then suddenly jumps ahead, while for $\theta=80\dgr$ the shock propagation is significantly smoother \citep[see also][]{Lee+04}. 
The shock reformation is revealed in Figure \ref{fig:trans}, which shows the evolution of the $x-p_x$ phase space for $\theta=0$.
At $t=166.5\omega_c^{-1}$ there is a sharp transition between the cold upstream beam and the isotropic downstream distribution, which is associated with compression and pressure increase, and in turn with an electric field $E_x\propto-\nabla P_e$ directed upstream;
the electron pressure $P_e\propto n^{\gamma_e}$ is taken as polytropic, with an effective adiabatic index satisfying the shock jump conditions with thermal equilibration between downstream ions and electrons ($\gamma_e\approx 4.3$ for $M=20$)\footnote{We checked that this parametrization of the electron physics has little effect on the overall shock structure and on ion injection \citep[see also][]{leroy+82}.}. 
Ions impinging on such a barrier are specularly reflected back into the upstream (beam with positive $p_x$ at $x\gtrsim 900c/\omega_p$ in Figure \ref{fig:trans}).
Reflected ions gyrate into the upstream, eventually producing a new discontinuity about one gyroradius $v_{\rm sh}/\omega_c\sim 20c/\omega_p$ ahead of the old one.
Meanwhile, the first barrier is smoothed out, and ions trapped between the two discontinuities  are rapidly isotropized.
This indicates that a necessary condition for an ion to be energized is to impinge on the shock when the potential barrier is at its maximum;
therefore, the  shock reformation timescale ($\sim \pi/\omega_c$, half the gyration time) sets the duty cycle for ion injection. 
 
\begin{figure}\centering
\includegraphics[trim=10px 150px 0px 160px, clip=true, width=.485\textwidth]{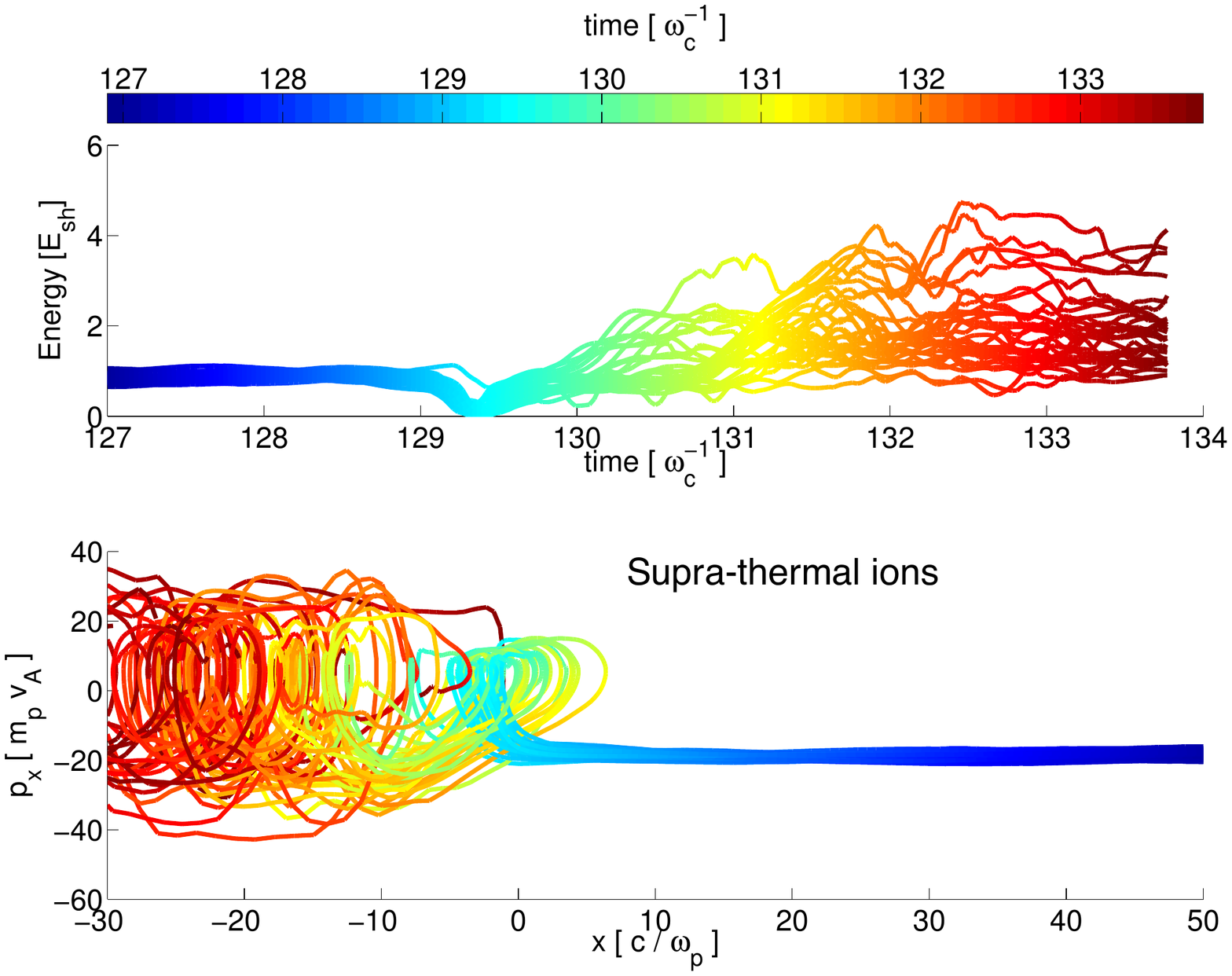}
\includegraphics[trim=10px 160px 0px 160px, clip=true, width=.485\textwidth]{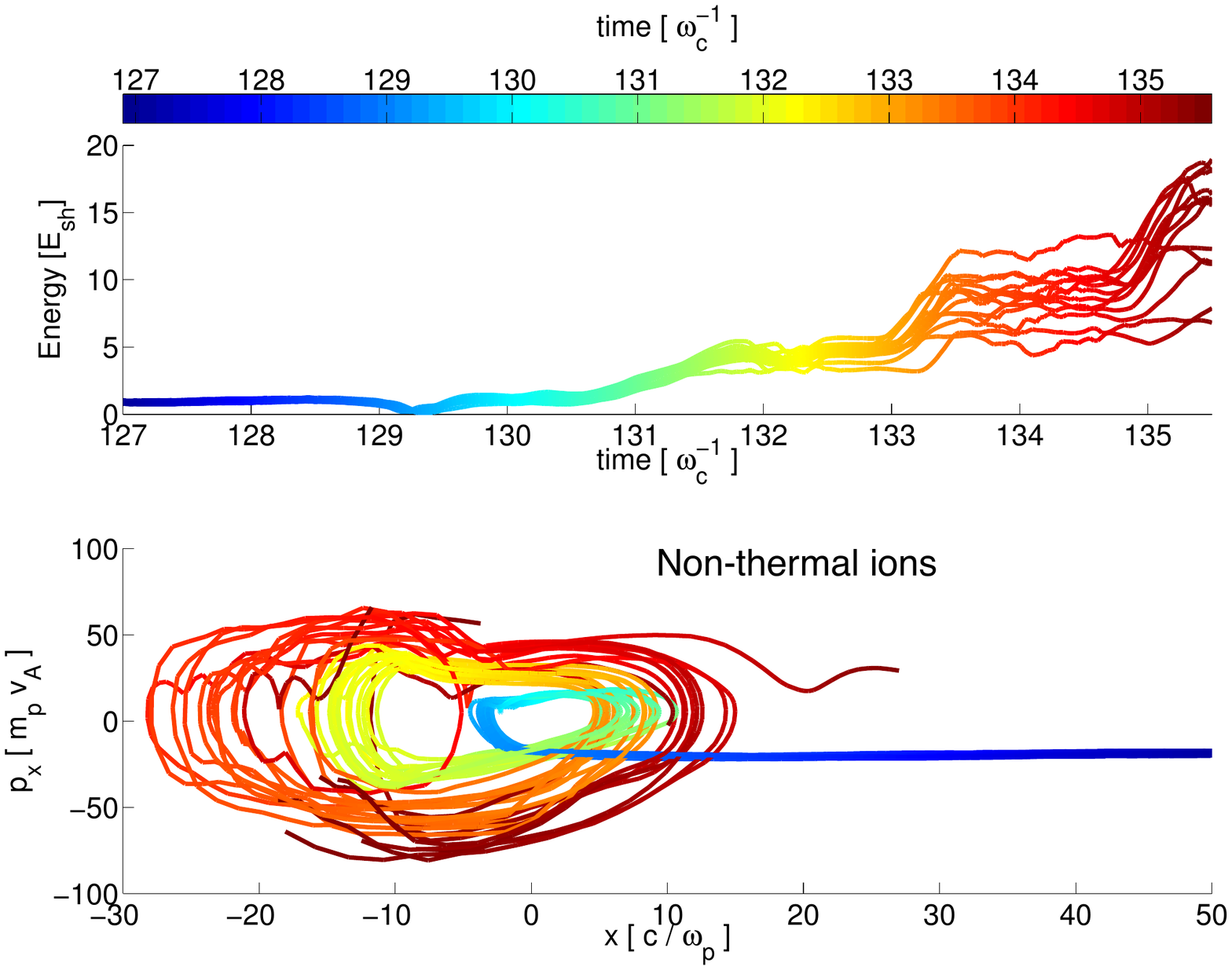}
\caption{\label{fig:part} Energy evolution and $x-p_x$ phase space distribution for typical supra-thermal (top panels) and non-thermal ions (bottom panels), for a parallel shock with $M=20$.
The $x$-axis is rescaled such that the shock is at $x(t\approx 130\omega_c^{-1})=0$, and moves with constant velocity.
}
\end{figure}

By tracking the trajectories of individual ions, we find that all the ions that eventually achieve energies larger than $E_{\rm sh}$ are reflected by the potential barrier at their first shock encounter. 
After reflection, ions perform a few gyrations around the shock, gaining energy via shock drift acceleration \citep[SDA, see, e.g.,][]{scholer90,Su+12}.
At any shock reformation, $\sim25\%$ of the incoming ions are reflected, but not all of them enter DSA.  
More precisely, ions impinging on the shock turn into: 
\begin{itemize}
\item \emph{thermal ions}, which encounter a barrier too weak to reflect them, and immediately cross downstream;
\item \emph{supra-thermal ions}, which are specularly reflected, and achieve $E\lesssim 6E_{\rm sh}$ via SDA before being advected downstream (Figure \ref{fig:part}, top panels);
\item \emph{non-thermal ions}, which are reflected, energized up to $E\gtrsim 10E_{\rm sh}$, and eventually escape toward upstream (Figure \ref{fig:part}, bottom panels). 
\end{itemize}
Non-thermal ions are injected into the DSA process: they later scatter on self-generated magnetic fluctuations and diffuse back to the shock for further energization.
The existence of supra-thermal ions, instead, demonstrates that reflection is necessary but not sufficient for DSA injection\footnote{Note that supra-thermal ions are present also at oblique shocks, which do not show DSA tails (Paper I).}.

Several authors have already pointed out the importance of shock reformation \cite[e.g.,][and references therein]{Lee+04, Su+12} and specular reflection \cite[e.g.,][]{gosling+82,BS84,scholer-terasawa90,gg13} for the production of energetic ions \cite[see also][]{leroy-winske83,ks91,giacalone+92}.
However, the distinction between supra-thermal and non-thermal ions (which undergo a similar initial reflection, but have different fates), has not been fully characterized.
Our goal is to develop a theory able to predict the spectrum and fractions of thermal, supra-thermal, and non-thermal ions observed in simulations, as a function of shock strength and inclination. 

\section{Injection Momentum}\label{sec:inj}
We adopt the formalism developed by \cite{STG83} and \cite{BS84} for studying ion reflection off the shock discontinuity, with the important difference that we account for specular reflection in the downstream frame (hereafter, DSF) rather than in the shock frame, since the potential barrier stalls before reformation \citep[see also][]{Lee+04}.
We introduce the \emph{de Hoffmann--Teller} frame (hereafter, HTF)\footnote{The HTF is defined only in a time-averaged sense for reforming shocks, and does not exist for $\vartheta\gtrsim\arccos(v_{\rm sh}/c)$.}, in which the shock is at rest and there is no motional electric field \citep{HT50}, and the ortho-normal triad $({\vb},{\vz},{\vx})$, such that ${\bf B}_0\equiv B_0{\vb}$ and the shock normal ${\vn}\equiv(\cth,\sth,0)$.
The velocity of the HTF with respect to the DSF is
\begin{equation}\label{eq:wHT}
\vw_{\rm HT}=\frac{1}{\cth}\vb+\left(\frac{1}{r}-1\right)\vn,
\end{equation}
where $r$ is the shock compression ratio.
In this section, we normalize velocities to the shock velocity in the upstream frame, $V_{\rm sh}=(1+1/r)v_{\rm sh}\simeq 1.25 v_{\rm sh}$, and indicate with $\vv$ ($\vw$) velocities in the DSF (HTF);
in general:
\begin{equation}\label{eq:HT}
\vv=\vw+\vw_{\rm HT}.
\end{equation}
In the HTF, any velocity can be decomposed as
\begin{equation}\label{eq:wt}
\vw(t) = \vp \vb + \vg[\cos(\tau+\phi)\vz-\sin(\tau+\phi)\vx],
\end{equation}
where $\tau\equiv t\omega_c$, $\vp$ is the guiding center velocity, $\vg$ is the gyro-speed, and $\phi$ is the gyro-phase that gives the correct ${\bf v}(t=0)$. 
A velocity written in the DSF as
\begin{equation}\label{eq:vn}
\vv=v_n\vn+\delta\vv,
\end{equation}
can be rewritten in the HTF as (Equations (\ref{eq:wHT})--(\ref{eq:wt})):
\begin{align}
\begin{split}
w_n &= \vw\cdot\vn = v_n-1/r;\\
\vp &= (w_n+1)\cth-1/\cth+\delta v_{\parallel};\\
\vg &= \sqrt{[(w_n+1)\sth+\delta v_{\zeta}]^2+\delta v_{\xi}^2};\\
\phi&= \arcsin(\delta v_{\xi}/w_g).
\end{split}
\end{align}
If an ion undergoes a specular reflection in the DSF ($v_n\to-v_n$), its final velocity in the HTF reads:
\begin{align}\label{eq:wR}
\begin{split}
w_{R,\parallel} &= \vp-2v_n\cth;\\
w_{R,g} &= \sqrt{(w_g\cos\phi-2v_n\sth)^2+(w_g\sin\phi)^2};\\
\phi_R&= \arcsin(w_g\sin\phi/w_{R,g}).
\end{split}
\end{align}
If $w_{R,\parallel}<0$, the ion guiding center motion is towards downstream, and the ion is advected away.
If $w_{R,\parallel}>0$, the guiding center motion is upstream, but gyration may still bring it back to the shock.
By integrating Equation (\ref{eq:wt}), with $\vw_R$ given by Equation (\ref{eq:wR}), we obtain the ion's displacement along the shock normal:
\begin{equation}\label{eq:Xt}
 X_n(\tau)\omega_c=w_{R,\parallel} \tau \cth+w_{R,g}\sth(\sin\tau-\sin\phi_R).
\end{equation}
For given $\vartheta$ and $\vw_R$, if there exists a $\tau_*$ such that  $X_n(\tau^*)=0$, a reflected ion reencounters the shock, impinging with normal velocity
\begin{equation}\label{eq:wn}
w_n(\tau_*)=w_{R,\parallel}\cth+w_{R,g}\cos(\tau^*+\phi_R).
\end{equation}
We define the characteristic loss angle $\thl$ as the smallest angle for which $X_n(\tau)=0$ has a solution, so that, if $\vartheta<\thl$, reflected ions escape upstream and are injected into DSA.

Let us consider a cold upstream ion with $v_n=-1+1/r$, and $\delta\vv=0$, which corresponds to
\begin{equation}
w_{R,\parallel}=2\left(1-\frac{1}{r}\right)\cth-\frac{1}{\cth}.
\end{equation}
For cold ions $\thl\approx 31.6\dgr$ (accounting for a finite-temperature beam only introduces a small spread in $\thl$, $\lesssim 5\dgr$ for $M_s=4$), which means that ions can be injected by a single reflection only at shocks with $\theta\lesssim 35\dgr$.
However, quasi-parallel shocks isotropize and amplify the pre-shock magnetic field (Paper II), which implies that the effective shock inclination is typically $\theta\approx 45\dgr$;
therefore, injection at (initially) quasi-parallel shocks cannot rely just on simple reflection of thermal upstream ions.
If $\vartheta\geq\thl$  the ion reencounters the shock, and may either penetrate downstream, or be reflected again, depending on the ion ``normal'' energy $\frac m2 w_n(\tau_*)^2V_{\rm sh}^2$ being sufficient to  overcome the shock potential $\Delta\Phi$, estimated as
\begin{align}
E_x &\simeq-\frac{\Delta \Phi}{\Delta x}\simeq-\frac{1}{ne}\frac{\Delta P_e}{\Delta x}\to\\
e\Delta\Phi &\simeq\frac{\Delta P_e}{n}\simeq \frac{P_{e,ds}}{n}\simeq \left(1-\frac{1}{r}\right)\frac{m V_{\rm sh}^2}{2},
\end{align}
if downstream electrons are in equipartition with ions; the last equality comes from jump conditions for strong shocks.
Ions penetrate the shock barrier if
\begin{equation}\label{eq:cond}
w_n(\tau_*)<-\sqrt{\Psi}; \quad \Psi\equiv\frac{r-1}{r}\frac{2e\Delta\Phi}{mV_{\rm sh}^2},
\end{equation}
with $\Psi=1$ for Rankine-Hugoniot conditions.
\begin{figure}\centering
\includegraphics[trim=37px 200px 60px 200px, clip=true, width=.485\textwidth]{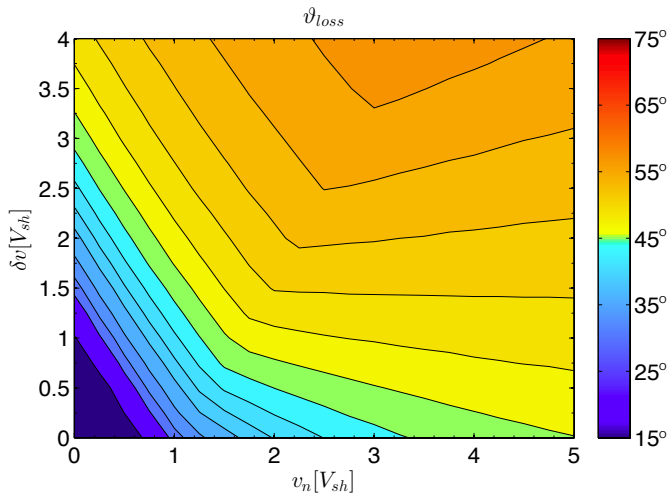}
\caption{\label{fig:thl} 
Maximum shock inclination allowing a reflected ion to escape upstream, as a function of the pre-reflection velocity, with $\delta \vv=(1,1/\sqrt{2},1/\sqrt{2})\delta v$.
The modulus of the minimum velocity necessary to escape from a shock with $\theta\sim 45\dgr$ (green contour) is typically $v_{\rm inj}\gtrsim 2.5-3.5V_{\rm sh},$ corresponding to $E_{\rm inj}\gtrsim 5-10 E_{\rm sh}$.}
\end{figure}
If condition (\ref{eq:cond}) is not satisfied, reflected ions reencounter the shock, being no longer cold because they have experienced SDA, and have been partially isotropized.

We can calculate the velocity that ions need for escaping upstream by looking at the phase space for which $X(\tau^*)=0$ has no solutions. 
Figure \ref{fig:thl} shows $\thl$ as a function of the pre-reflection ion velocity, decomposed as in Equation (\ref{eq:vn}). 
The green contour indicates the velocity components that permit reflected ions to escape from a DSA-efficient shock with $\theta\approx 45\dgr$;
the corresponding minimum injection velocity is $v_{\rm inj}\equiv\sqrt{v_n^2+\delta v^2}|_{\rm green} \gtrsim 2.5-3.5$, and the minimum injection energy is $E_{\rm inj}\gtrsim 5-10E_{\rm sh}$, in good agreement with simulations (Paper I);
different orientations of $\delta \vv$ return similar values.

Injection into DSA is suppressed for shocks with $\theta\gtrsim45\dgr$ because ions can escape very oblique shocks only with $v_{\rm inj}\gtrsim 4$ (Figure \ref{fig:thl}). 
The achievement of such velocities requires a few more SDA cycles, and, since at every cycle ions have a finite probability to pierce the shock barrier and to be lost downstream, the fraction of ions that can achieve larger velocities becomes increasingly smaller.
Our results differ from those by \cite{BS84}, who assumed reflection in the shock frame rather than in the DSF and found that for $\theta\lesssim 55\dgr$ all the reflected ions are injected.

\section{A Minimal Model for Ion Injection}\label{sec:suprath}
We now construct a minimal model that accounts for the observed: 1) periodic shock reformation, 2) fraction and trajectories of reflected ions, and 3) ion spectrum above thermal energies.
Since we are interested in DSA-efficient shocks, without loss of generality we consider $\theta\approx 45\dgr$, and $M$ of a few tens (magnetic field amplification effectively reduces $M_A$ and $M_s$ in the precursor of stronger shocks, see Paper II).
\begin{figure}\centering
\includegraphics[trim=68px 260px 62px 260px, clip=true, width=.485\textwidth]{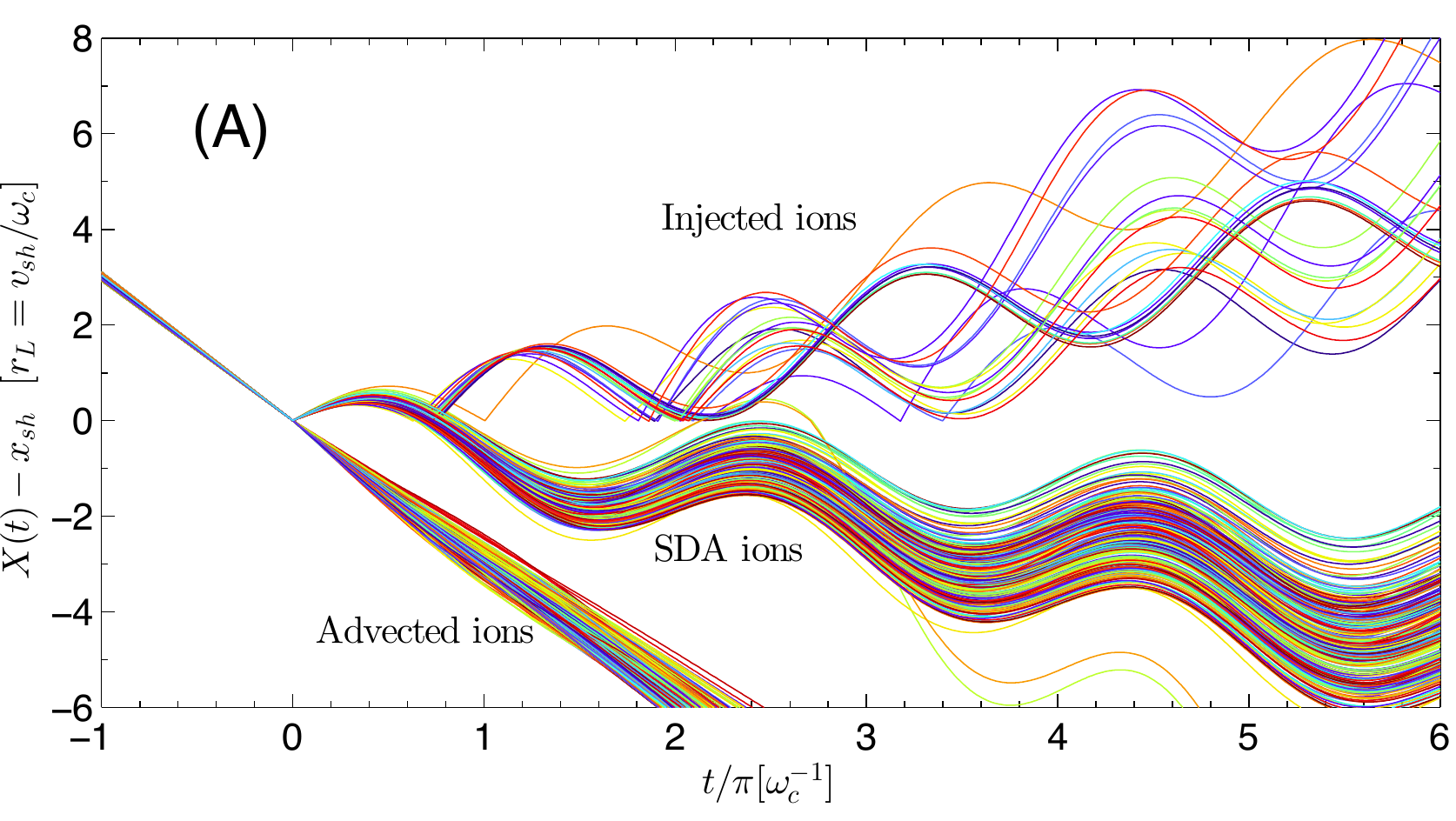}
\includegraphics[trim=12px 250px 10px 240px, clip=true, width=.485\textwidth]{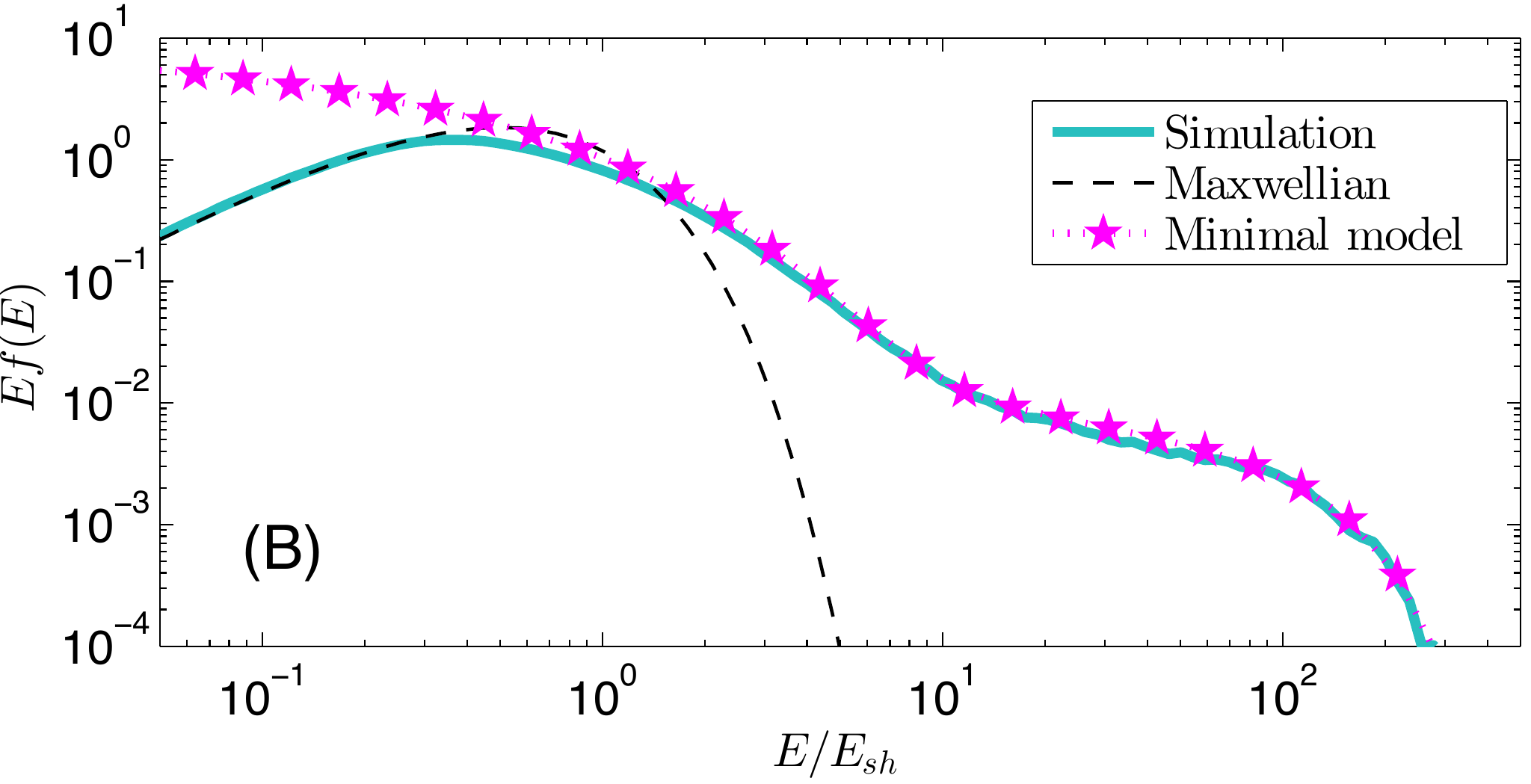}
\caption{\label{fig:traj} 
(A) Trajectories of test-particles impinging at random times on a periodically-reforming shock with $M=10$ and $\theta\simeq 45\dgr\pm 2\dgr$ (Section \ref{sec:suprath}); for each ion, $t=0$ corresponds to the first shock encounter.
Ions may either not reflect (``Advected ions''), or experience SDA before ending up downstream (``SDA ions''), or escape upstream after a few reflections (``Injected ions'').\\
(B) Post-shock ion spectrum for a parallel shock with $M=20$. 
Our minimal model (magenta symbols) perfectly matches the spectrum obtained in simulations (solid line).
}
\end{figure}

Inspired by simulations, we model the periodic shock reformation by imposing the potential barrier to spend $\sim 25\%$ of the time in a ``high'' state with $\Psi_{\rm os}\equiv n_{\rm os}/4\approx 7/4$, defined as the overdensity at the density peak (the ``overshoot'', see Figure \ref{fig:rho}), and the rest of the time in a ``low state'', with normalization chosen such that $\langle\Psi\rangle=1$ when averaged over a period\footnote{Note that the barrier's duty cycle is determined by the gyration of reflected ions, and not by the exact value of $\Psi_{\rm os}\gtrsim1$.}.

We calculate the trajectories of test-particle ions impinging on the shock at random times (Equation (\ref{eq:Xt})), performing a specular reflection (Equation (\ref{eq:wR})) whenever ions reencounter the shock with normal velocity too small to penetrate the barrier (Equation (\ref{eq:cond})).
Figure \ref{fig:traj}A shows the displacement from the barrier of several ions impinging at $t=0$ on a shock with $M=10$, and inclination $\theta=45\dgr\pm2\dgr$.
We recover the populations of Section \ref{sec:reform}, now labeled as: 1) ``advected ions'' ($\sim 75\%$ of the total), which impinge on the shock when the barrier is in the low state, and remain trapped downstream;
2) ``SDA ions'' ($\sim 20\%$), which end up in the downstream after having crossed the barrier once or twice, gaining a factor of a few in energy in the process;
3) ``injected ions'' ($\lesssim 4\%$), which escape upstream after two to four reflections, and would enter DSA in a full simulation in the presence of upstream scattering.

By using the approach put forward by \cite{bell78a}, we also calculate the ion spectrum in the supra-thermal region. 
If the fractional energy gain at each acceleration cycle is $\mathcal E\equiv E_{\rm fin}/E_{\rm in}-1$, with $E_{\rm in}(E_{\rm fin})$ the initial (final) energy, and the probability of leaving the acceleration region is $\mathcal P$, the expected particle spectrum reads 
\begin{equation}\label{eq:gam}
f(E)\propto E^{-1-\gamma}; \quad \gamma\equiv-\frac{\ln(1-\mathcal P)}{\ln(1+\mathcal E)}.
\end{equation}
If $\mathcal P,\mathcal E\ll1$, then $\gamma\simeq \mathcal P/\mathcal E$; 
for relativistic particles $\mathcal E\simeq\mathcal P\simeq V_{\rm sh}/c$, and one gets the universal DSA ion spectrum $f(E)\propto E^{-2}$, while for non-relativistic particles $\mathcal E\simeq 2V_{\rm sh}/v$ and $\mathcal P\simeq V_{\rm sh}/v$, so that $f(E)\propto E^{-1.5}$.

The energy gain $\mathcal E$ is independent of the acceleration mechanism (SDA or DSA).
Instead, the probability of leaving the acceleration region is insensitive to the shock discontinuity for DSA ions, but is regulated by the duty cycle of the potential barrier in the SDA regime.
For non-thermal ions with $E\gtrsim 10E_{\rm sh}$, we assume $\mathcal P_{\rm nt}\simeq V_{\rm sh}/v$ as in usual DSA theory, where $\mathcal P$ is determined by the advection of an isotropic ion distribution.
Simulations suggest $\mathcal P_{\rm st}\simeq 0.75$ (independent of energy) for supra-thermal ions, whose spectrum deviates from a power-law and is steeper than in the DSA region.
Figure \ref{fig:traj}B shows the ion spectrum obtained in a hybrid simulation of a parallel shock with $M=20$, compared with the spectrum obtained by using the full Equation (\ref{eq:gam}), and the prescriptions above for $\mathcal P$ and $\mathcal E$, plus a cut-off at $E_{\rm max}\simeq 180E_{\rm sh}$. 
Our minimal model remarkably reproduces the simulated spectrum, which deviates from a Maxwellian above $\sim 2E_{\rm sh}$, shows a steep ``bridge'' in the supra-thermal region, and matches the standard DSA prediction above $\sim 10E_{\rm sh}$.

The normalization of the non-thermal tail at DSA-efficient shocks is determined by the number $\mathcal N$ of SDA cycles  needed to accelerate ions above the injection energy for a shock with inclination $\theta\approx 45\dgr$.
With the procedure above, we calculate $\mathcal N\approx 2.4$, and an injection fraction of  $\eta\equiv(1-\mathcal P_{\rm st})^{\mathcal N}\sim 3.6\%$, in excellent agreement with simulations. 
Injection at shocks with $\theta\gtrsim 50\dgr$ is strongly suppressed because it requires higher $E_{\rm inj}\gtrsim 10E_{\rm sh}$, corresponding to $\mathcal N \gtrsim 4$, and at each SDA cycle ions have $\sim 75\%$ probability of being lost downstream; for instance, for $\theta=50\dgr$ we find $\mathcal N\approx 3.8$, and $\eta\sim 5\times 10^{-3}$.
This explains why DSA efficiency is almost constant for $\theta\lesssim 45\dgr$, and drops rapidly above $\theta \sim 50\dgr$ (Figure 3 in Paper I);
moreover, acceleration efficiency is almost independent of the shock strength for $M\gtrsim 10$, which suggests that our recipes hold for any strong shock.

\section{Discussion}\label{sec:disc}
Ion injection is often accounted for with a \emph{thermal leakage} model \citep[see, e.g.,][and references therein]{eje81,malkov98,KJG02,bgv05}; downstream thermal ions of sufficiently large energy ($\gtrsim E_{\rm inj}$) are assumed to be injected because their gyroradius encompasses the shock thickness, which is however not resolved in macroscopical approaches to DSA.
Monte Carlo models \citep[e.g.,][]{jones-ellison91} do not need to specify $E_{\rm inj}$, and can reproduce the supra-thermal bridge measured, e.g., at the Earth's bow shock \citep{EMP90}, but need an a priori parametrization of the ion mean free path \citep[see][for a comparison of different approaches to DSA]{comparison}.
Our self-consistent picture is intrinsically different, in that supra-thermal ions have never been thermalized, and their propagation is never diffusive.
The scheme outlined in Section \ref{sec:inj} provides a realistic description of the injection microphysics, as well as a simple parametrization of $E_{\rm inj}/E_{\rm sh}$ for phenomenological purposes. 

When shocks are not strong, three effects may become important: 
1) for $M\lesssim 4$ the overshoot vanishes \citep[][]{leroy+82}, and a larger fraction of ions is advected downstream;
2) $\mathcal E(r<4)$ is smaller, and reaching $E_{\rm inj}$ requires more SDA cycles;
3) magnetic field amplification is reduced (Paper II), and the effective shock inclination is $\theta\lesssim45\dgr$, which in principle helps injection (Figure \ref{fig:thl}).    
However, the net effect is that the energy fraction in DSA ions is lower for low-$M$ shocks (Figure 3 in Paper I).
This leads to a crucial feedback: when $\sim 10\%$ of the ram pressure is in CRs, the development of a shock precursor reduces $M_s$ at the subshock (Paper I, Section 6), which suppresses injection and prevents a more prominent shock modification.
In simulations we consistently see the normalization of the non-thermal spectrum to decrease with time (Figure 2 in Paper II), which keeps the energy in the non-thermal tail saturated at $\sim 10\%$ despite the increase of $E_{\rm max}\propto t$ (Paper III).

We stress that the dependence of ion injection on the shock obliquity is not an artifact of 2D simulations (see 3D runs in Paper I).
At real quasi-parallel shocks \citep[e.g.,][]{burgess+05}, reformation is unlikely coherent along the shock surface; also, long-wavelength waves produced by CR instabilities perturb the shock front \citep[][]{filam}, interfering with its natural cyclotron period. 
Pre-existing turbulence may locally affect the shock inclination, possibly providing patchy ion injection also for shocks with a globally quasi-perpendicular magnetic field \citep{Giacalone05}.
Nevertheless, our results are still expected to be valid in a time/space-averaged sense, as injection is a local process.  

In this work we did not account for injection of electrons, and of heavy ions, which are preferentially accelerated in Galactic CRs;  
we defer the generalization of the presented formalism to particles with different mass/charge ratios to forthcoming publications.

\section{Conclusions}\label{sec:conclusions}
We investigated ion injection in non-relativistic collisionless shocks with kinetic hybrid simulations, in which shock structure and ion distribution are calculated self-consistently. 
We focused on DSA-efficient quasi-parallel shocks, and attested to their periodic reformation due to the collective reflection of ions off the shock potential barrier.
Because of such a time-dependent barrier, on average $\sim25\%$ of the ions impinging on the shock are reflected and energized via SDA;
nevertheless, not all of the reflected ions gain enough energy to enter DSA.
For the effective magnetic inclination of DSA-efficient shocks ($\theta\sim45\dgr$), reflected ions must undergo two to three gyrations (SDA cycles) around the shock before escaping upstream (Section \ref{sec:suprath});
since at each cycle $\sim 75\%$ of them are trapped downstream of the oscillatory barrier, only $\lesssim4\%$ of the incoming ions survive after several SDA cycles to be injected into DSA.

We presented a formalism for studying supra-thermal ions in their multiple reflections, and calculated the minimum energy ions need to escape upstream of the shock, and enter DSA (Figure \ref{fig:thl}).
We also explained the observed dependence of the injection fraction on the shock inclination (Paper I), providing a general explanation for the reason why DSA is most prominent at quasi-parallel shocks.
With our minimal shock model, spectrum and normalization of the ion spectra obtained in simulations are well reproduced.
Our findings provide a theory of ion injection that is of primary importance for understanding ion acceleration in interplanetary shocks, and in several astrophysical objects, such as SNRs and clusters of galaxies.

\vspace{-2mm}
\subsection*{}
We thank L.\ Gargat\'e for providing \emph{dHybrid} and the referee for precious comments. 
This research was supported by NASA (grant NNX14AQ34G to DC), and facilitated by the Max-Planck/Princeton Center for Plasma Physics and by the Simons Foundation (grant 267233 to AS).
Simulations were performed on the computational resources provided by the Princeton High-Performance Computing Center, by NERSC (supported by the Office of Science of the U.S. Department of Energy under Contract No.\ DE-AC02-05CH11231), and by XSEDE (allocation No.\ TG-AST100035).

\end{document}